\documentclass[preprint,showpacs,preprintnumbers,amsmath,amssymb,pre]{revtex4}

\usepackage{graphicx}
\usepackage{dcolumn}
\usepackage{bm}

\begin{document}

\preprint{}

\title{Collective and static properties of model two-component plasmas}

\author{Yu.V. Arkhipov$^{1}$, A. Askaruly$^{1}$, D. Ballester$^{2}$, A.E.
Davletov$^{1}$, G.M. Meirkanova$^{1}$, and I.M. Tkachenko$^{2}$ }
\affiliation{$^{1}$Department of Optics and Plasma Physics, al-Farabi
Kazakh National University, Tole Bi 96, Almaty 050012, Kazakhstan, }
\affiliation{$^{2}$Department of Applied Mathematics, Polytechnic University of
Valencia, Camino de Vera s/n, 46022 Valencia, Spain}

\date{\today}

\begin{abstract}
Classical MD data on the charge-charge dynamic structure factor of
two-component plasmas (TCP) modelled in Ref. \cite{H2} are analyzed
using the sum rules and other exact relations. The convergent power moments of
the imaginary part of the model system dielectric function are expressed in
terms of its partial static structure factors, which are computed by
the method of hypernetted-chains using the Deutsch effective potential.
High-frequency asymptotic behavior of the dielectric function is specified to
include the effects of inverse bremsstrahlung. The agreement with the MD data
is improved, and important statistical characteristics of the model TCP, such as
the probability to find both electron and ion at one point, are determined.

\end{abstract}

\maketitle

\section{Introduction}

Classical molecular dynamics (MD) simulations of model two-component plasmas
(TCP) were carried out in Ref. \cite{H2} more than twenty five years ago. The system modelled
in this pioneer work was a fully ionized strongly coupled hydrogen TCP of
temperature $T$ consisting of electrons ($a=e,$ $Z_{e}=-1$) with the number
density $n$ and protons ($a=i,$ $Z_{i}=+1$) with the same number density. The
dynamic and static characteristics studied were the charge-charge dynamic
structure factor, the species partial static structure factors and the static
radial distribution functions, etc. Classical statistical averages were
computed on the basis of the ergodic hypothesis while the quantum delocalization
preventing the collapse were taken into account through the use of the Deutsch
pair effective potential \cite{D} arising from the quantum-diffraction effects without including the exchange or symmetry contribution,
\begin{align}
\varphi_{ab}\left(  r\right)   &  =Z_{a}Z_{b}\frac{e^{2}}{r}\left(
1-\exp\left(  -\kappa_{ab}r\right)  \right)  ,\quad a,b=e,i , \label{Dpot}\\
\kappa_{ab}  &  =\sqrt{\frac{2\pi\mu_{ab}}{\beta\hbar^{2}}}, \quad\mu
_{ab}=\left(  m_{a}^{-1}+m_{b}^{-1}\right)  ^{-1},\nonumber
\end{align}
where $\mu_{ab}$ is the reduced mass of an $a-b$ pair and $\beta^{-1}$ is the temperature in energy units. The fact that the potentials (\ref{Dpot}) remain finite
as $r\rightarrow0$ is a consequence of the uncertainty principle and prevents
the collapse to which we have already referred. In the temperature range of
interest
\[
\kappa_{ii}a\gg1,\quad a=\sqrt[3]{\frac{3}{4\pi n}}%
\]
being the Wigner-Seitz or the "ion sphere" radius. Thus the effective ion-ion
interaction is virtually identical to the bare Coulomb potential at all separations.

In addition, in Ref. \cite{H2} the potential (\ref{Dpot}) was also employed to determine the static
properties in the hypernetted-chain (HNC) approximation.

Since 1981, little effort was made to simulate this benchmark high-energy
density system and study its static and dynamic properties. We mention a few
related works: (i) the static electrical conductivity was studied in Refs.
\cite{hedw} and \cite{TF}, (ii) the TCP dynamic characteristics in a
different range of values of the wavenumber were investigated in Ref. \cite{ZP}.

The results of Ref. \cite{H2} were analyzed in Ref. \cite{ATMV} using the sum rules and
other exact relations. An overall agreement with the MD results was obtained
in Ref. \cite{ATMV}, where the frequency moments of the imaginary part of the
plasma inverse dielectric function, $\epsilon^{-1}(k,\omega)$, (the sum
rules) were calculated for the bare Coulomb potential.

Our aim here is to reexamine the simulation data of Ref. \cite{H2} and the
theoretical results of Ref. \cite{ATMV} within the moment approach and using the
method of effective potentials to evaluate the static characteristics of the TCP.

Mathematical details of the moment approach are provided in the following Section. A model Nevanlinna parameter function taking into account the fractional asymptotic form of the imaginary part of the system dielectric function is suggested in Sect. \ref{Q}; in Sect. \ref{M} the convergent power moments of the loss function $\left[
-\operatorname{Im}\epsilon^{-1}(k,\omega)/\omega\right]  $ of the model system are calculated using the Deutsch effective potential and are computed by the method of hypernetted-chains using the same potential; for a recent review of the method of effective potentials see Ref. \cite{E}. Numerical results and conclusions are presented in Sects. III and IV, respectively. The agreement with the MD data is
improved significantly, and, simultaneously, important statistical characteristics of the
model TCP in concern, like the probability to find both electron and ion at
one point $\left[  g_{ei}\left(  0\right)  \right]  $, are determined, direct
and exchange interaction contributions being compared.

\section{The method of moments}

\subsection{The background}

The MD results of Ref. \cite{H2} on the charge-charge dynamic structure factor
$S_{zz}(k,\omega)$ were modelled in Ref. \cite{ATMV} using the moment approach
which automatically takes into account the sum rules and other exact relations.

The starting point in the application of the method of moments \cite{AT} to
the calculation of the system dynamic correlation function is the
fluctuation-dissipation theorem (FDT) which relates the latter to the system
dissipation characteristic, the Green function, whose power frequency moments,
by virtue of the Kubo theory of linear response, can be directly expressed in
terms of the static correlators of the time derivatives of the system
observables. These static characteristics can be expressed, using the system model Hamiltonian, in terms of
the system structural static correlation functions like the radial
distribution functions or the static structure factors, and this is the only
step of our approach where the system model interferes. Otherwise the
relations we use are model-free. They are based on mathematical results which
are independent of the physical details, i.e., the interaction potential. This permits to apply the moment
approach to non-perturbative systems which lack small parameters, such as the one
we deal with here.

Indeed, in a classical Coulomb system the characteristic perturbation
parameter is the potential energy of two charges at an average distance, i.e.,
the Wigner-Seitz radius $a$, divided by the characteristic kinetic energy or the system temperature:
\[
\Gamma=\beta e^{2}/a .
\]
The achievement of the work by
Hansen and McDonald \cite{H2} was that the values of $\Gamma\gtrsim1$
(precisely, $\Gamma=0.5$ and $\Gamma=2$) and $r_{s}=0.4$ and $r_{s}=1,$ where
\[
r_{s}=\frac{a}{a_{0}}=\frac{ame^{2}}{\hbar^{2}}%
\]
$a_{0}$ being the Bohr radius, were considered, which correspond to
strongly coupled Coulomb systems or high-energy density (HED) matter. Under
these conditions the Landau length, the Wigner-Seitz radius,
and the Debye radius
\[
\lambda_{D}=\frac{a}{\sqrt{3\Gamma}}%
\]
are of the same order of magnitude, so that no effective screening takes place:
there is only one or less particles in the Debye sphere. Hence, standard
theoretical treatments, e.g., the kinetic theory are not applicable, and we
need alternative approaches such as the one based on the method of moments.

There is an additional dimensionless parameter which is commonly used to characterize the kind of statistical systems we consider. This is the so-called degeneracy parameter,
\[
\theta^{-1}=\beta E_{F}=1.84159\frac{\Gamma}{r_{s}}
\]
which compares the Fermi energy of the system, $E_{F}$, to its thermal energy. For the thermodynamic conditions considered in Ref. \cite{H2} and here $\theta^{-1}\gtrsim 1$, which means that strongly coupled two-component plasmas are intrinsically quantum statistical systems, and we might expect their physical properties to be greatly influenced by their quantum mechanical nature.

Nevertheless, due to the significant difficulties encountered in the analytical and computational modelling of these systems, it is a commonplace to use classical statistical approaches to investigate multicomponent plasmas. These classical approximations, like MD or HNC calculations, do require the application of potentials which effectively take into account the quantum mechanical nature of these systems, especially the quantum diffraction preventing the classical Coulomb collapse.

As we have said, the only step where the physical model under consideration appears explicitly in the method of moments is the modelling of the system Hamiltonian required to compute the frequency moments of the Green function (see Sect. \ref{M}). Strictly speaking, these calculations can be carried out according to the rules of quantum statistical mechanics, and therefore no effective potential is needed. This means that the potential appearing in the Hamiltonian must be understood as the real interaction among charged particles at the microscopic level, which is, of course, the Coulomb potential.

However, our main aim here is to compare the static and dynamic characteristics obtained for the TCP with those of Ref. \cite{H2}. Thus, we will work under the same general framework in order to better assess the applicability of our approach. In this sense, the usage of the classical version of the FDT is justified:
\begin{equation}
S_{zz}\left(  k,\omega\right)  =\frac{\mathcal{L}\left(  k,\omega\right)
}{\pi\beta\phi\left(  k\right)  }, \label{Szz}%
\end{equation}
where
\[
\phi\left(  k\right)  =\frac{4\pi e^{2}}{k^{2}} ,
\]
\begin{equation}
\mathcal{L}\left(  k,\omega\right)  =-\frac{\operatorname{Im}\epsilon
^{-1}\left(  k,\omega\right)  }{\omega} \label{L}%
\end{equation}%
being the loss function of the system, which is assumed to be isotropic. The sum rules we employ are actually the power frequency moments of this loss
function
\begin{equation}
C_{\nu}\left(  k\right)  =\frac{1}{\pi}%
{\displaystyle\int_{-\infty}^{\infty}}
\omega^{\nu}\mathcal{L}\left(  k,\omega\right)  d\omega,  \quad \nu=0,2,4 .
\label{Cnu}%
\end{equation}
Notice that odd-order moments vanish due to the symmetry of the loss function.

The Nevanlinna formula of the classical theory of moments \cite{Akh1,Kr1} (for
a recent review see Ref. \cite{AK}) expresses the response function \cite{AT}%
\begin{equation}
\epsilon^{-1}(k,z)=1+\frac{\omega_{p}^{2}(z+Q)}{z(z^{2}-\omega_{2}%
^{2})+Q(z^{2}-\omega_{1}^{2})} \label{eps}%
\end{equation}
in terms of a Nevanlinna class function $Q=Q(k,z),$ analytic in the upper
half-plane $\operatorname{Im}z>0$ and possessing there a positive imaginary
part: $\operatorname{Im}Q(k, \omega+i\eta)>0,$ $\eta>0$. The function $Q(k,z)$
should also satisfy the limiting condition:
\begin{equation}
\frac{Q(k,z)}{z}\underset{z\uparrow\infty}{\rightarrow}0,\quad\operatorname{Im}z>0. \label{q}%
\end{equation}

Any such function admits the integral representation \cite{Akh1,Kr1}
\begin{equation}
Q(k,z)=ih\left(  k\right)  +%
{\displaystyle\int_{-\infty}^{\infty}}
\left(  \frac{1}{t-z}-\frac{t}{1+t^{2}}\right)  dg\left(  t\right)  \label{qg}%
\end{equation}
with $\operatorname{Re}h\left(  k\right)  \geq0$ and some non-decreasing
bounded function $g\left(  t\right)  $ such that%
\[%
{\displaystyle\int_{-\infty}^{\infty}}
\frac{dg\left(  t\right)  }{1+t^{2}}<\infty .
\]

The frequencies $\omega_{1}(k)$ and $\omega_{2}(k)$ in Eq. (\ref{eps}) are
defined by the respective ratios of the moments $C_{\nu}$ \cite{AT}
\begin{equation}
\omega_{1}^{2}=\omega_{1}^{2}\left(  k\right)  =C_{2}\left(  k\right)
/C_{0}\left(  k\right)  ,\quad\omega_{2}^{2}=\omega_{2}^{2}\left(  k\right)
=C_{4}\left(  k\right)  /C_{2}\left(  k\right)  , \label{omegi}%
\end{equation}
the latter are expressible in terms of the system static characteristics, see
Sect. \ref{M}.

It is easily seen \cite{Akh1} that the analytic prolongation of the loss
function onto the upper half-plane $\operatorname{Im}z>0$ constructed by means
of the Cauchy integral formula,
\[
\mathcal{L}(k,z)=\frac{1}{\pi}%
{\displaystyle\int_{-\infty}^{\infty}}
\frac{\mathcal{L}(k,\omega)}{\omega-z} ,
\]
admits the asymptotic expansion
\begin{equation}
\mathcal{L}(k,z\rightarrow\infty)\simeq-\frac{C_{0}\left(  k\right)  }%
{z}-\frac{C_{2}\left(  k\right)  }{z^{3}}-\frac{C_{4}\left(  k\right)  }%
{z^{5}}-\mathrm{o}\left(  \frac{1}{z^{5}}\right)  , \label{Lx}%
\end{equation}
while the expansion for the inverse dielectric function due to the
Kramers-Kronig relations,
\begin{equation}
\epsilon^{-1}(k,z)=1+\frac{1}{\pi}%
{\displaystyle\int_{-\infty}^{\infty}}
\frac{\operatorname{Im}\epsilon^{-1}\left(  k,\omega\right)  }{\omega
-z}d\omega , \label{KK}%
\end{equation}
reads:%
\begin{align}
\epsilon^{-1}(k,z  &  \rightarrow\infty)\simeq1+\frac{C_{2}\left(  k\right)
}{z^{2}}+\frac{C_{4}\left(  k\right)  }{z^{4}}+\mathrm{o}\left(  \frac
{1}{z^{4}}\right) \nonumber\\
&  \simeq1+\frac{\omega_{p}^{2}}{z^{2}}+\frac{\omega_{p}^{2}\omega_{2}^{2}%
}{z^{4}}+\mathrm{o}\left(  \frac{1}{z^{4}}\right)  ,\quad\operatorname{Im}%
z>0 . \label{epsx}%
\end{align}
Certainly, the expansions (\ref{Lx}) and (\ref{epsx}) stem from (\ref{eps}) by virtue of the condition (\ref{q})
\footnote[1]{In non-hydrogen-like plasmas with polarizable ions, the first
term on the r.h.s. of (\ref{eps}) and (\ref{epsx}) should be substituted by
the asymptotic value $\epsilon^{-1}\left(  k,\infty\right)  $, i.e., by the
value of $\epsilon^{-1}\left(  k,\omega\right)  $ with $\omega_{p}\ll\omega
\ll\omega_{x}$, $\omega_{x}$ being the characteristic excitation frequency of
ions and, possibly, atoms. Then $C_{0}\left(  k\right)  =\epsilon^{-1}\left(
k,\infty\right)  -\epsilon^{-1}\left(  k,0\right)  .$}; the convergence of the
moment $C_{0}\left(  k\right)  $ follows from the existence of the static
inverse dielectric function
\begin{equation}
\epsilon^{-1}(k,0)=\lim_{\eta\downarrow0}\epsilon^{-1}(k,i\eta) = 1 +P.V. \int_{-\infty}^{\infty}
\frac{{\rm Im}\epsilon^{-1}\left(k,\omega\right)} {\pi \omega}d\omega. \label{compress}
\end{equation}

It is important that the inverse dielectric function (\ref{eps}) satisfies the sum rules (\ref{Cnu}) by construction, irrespectively of the form of the Nevanlinna parameter function $Q(k,z)$ (see Sect. \ref{Q}). In particular, this means that the asymptotic expansions (\ref{Lx}) and (\ref{epsx}) are also valid for any function $Q(k,z)$.

Moreover, the inclusion of the moment $C_{0}(k)$ guarantees that the inverse dielectric function (\ref{eps}) meets the compressibility sum rule \cite{Ich1} whenever $\epsilon^{-1}(k,0)$ does. Within the present approach, this can be fulfilled by means of the relation between the charge-charge static
structure factor and the static value of the plasma dielectric function (\ref{compress}), see below in Sect. \ref{Stchar}, Eqs. (\ref{SzzFDT}) and (\ref{C0FDT}).

In general, there is no phenomenological basis for the choice of a unique $Q(k,z),$ which
would provide the exact expression for the loss function. The simplest is to
approximate $Q(k,z)$ by its static value $ih\left(  k\right)  $ directly
related to the static value $S_{zz}(k,0)$ of the dynamic structure factor
through Eq. (\ref{Szz}),
\begin{equation}
h(k)=\frac{\omega_{p}^{2}}{\pi\beta\phi\left(  k\right)  }\frac{\omega_{2}%
^{2}-\omega_{1}^{2}}{S_{zz}(k,0)\omega_{1}^{4}}. \label{h}%
\end{equation}
In this case the dynamic structure factor reads as
\begin{equation}
S_{zz}(k,\omega)=\frac{\omega_{p}^{2}}{\pi\beta\phi\left(  k\right)  }%
\frac{h\left(  \omega_{2}^{2}-\omega_{1}^{2}\right)  }{\omega^{2}(\omega
^{2}-\omega_{2}^{2})^{2}+h^{2}(\omega^{2}-\omega_{1}^{2})^{2}}. \label{Sh}%
\end{equation}
This is the approximation which was used as a basis of the analysis carried
out in Ref. \cite{ATMV}. If we use it now combined with more precise values for the
power moments calculated using the Deutsch potential \cite{D}, we fail to
predict the values of the Langmuir collective mode frequency, i.e., the
position of the lateral peak of the dynamic structure factor, and its width, i.e., the approximation $Q(k,z)=ih\left(  k\right)  $ is insufficient.

\subsection{The Nevanlinna parameter function \label{Q}}

It is quite clear from Eqs. (\ref{Szz}), (\ref{q}), and (\ref{eps}) that the parameter
function $Q(k,z)$ modifies both the position and the width of the Langmuir
line in the spectrum of collective excitations of the system, as reflected in
its charge-charge dynamic structure factor.

If we neglect the processes of energy absorption completely, we should put in
Eq. (\ref{Sh}) $h(k)=0^{+}$. Mathematically this means the utilization,
instead of the Nevanlinna formula (\ref{eps}), of the canonical solution of
the (truncated Hamburger) moment problem consisting of finding a loss function which satisfies the moment conditions (\ref{Cnu}) or,
equivalently, admits the expansion (\ref{Lx}).

The canonical solution physically corresponds to the existence in the system
of the diffusion (unshifted) and the infinitesimally-decaying Langmuir
(shifted) mode, and implies a Feynman-like decay-free approximation for the charge-charge dynamic structure factor
(\ref{Szz}):
\begin{equation}
S_{zz}(k,\omega)=\frac{C_{0}\left(  k\right)  }{\beta\phi\left(  k\right)
}\left\{  \frac{\omega_{2}^{2}-\omega_{1}^{2}}{\omega_{2}^{2}}\delta\left(
\omega\right)  +\frac{\omega_{1}^{2}}{2\omega_{2}^{2}}\left[  \delta\left(
\omega-\omega_{2}\right)  +\delta\left(  \omega+\omega_{2}\right)  \right]
\right\}  . \label{can}%
\end{equation}

To specify the Nevanlinna parameter function $Q(k,z)$ and, hence, the
non-canonical solution given by the Nevanlinna formula (\ref{eps}), we have to
reconsider the details of energy absorption in the system without violating
the sum rules.

Precisely, in addition to the expansion (\ref{epsx}), we want to satisfy the
well-known Perel'-Eliashberg asymptotic form for the imaginary part
of the dielectric function. The latter result can be summarized in the
following way.

In a completely ionized plasma for $\omega\gg\left(  \beta\hbar\right)  ^{-1}$
the microscopic acts of the electromagnetic field energy absorption become the
processes which are inverse with respect to the bremsstrahlung during pair collisions of
charged particles.

As it was shown by L. Ginzburg \cite{G}, this circumstance permits to use the
detailed equilibrium principle to express the imaginary part of the dielectric
function, $\operatorname{Im}\epsilon\left(  k,\omega\right)  $, of a
completely ionized plasma in terms of the bremsstrahlung cross section and
leads to the following asymptotic form of $\operatorname{Im}\epsilon\left(
k,\omega\right)  $ in a completely ionized (for simplicity, hydrogen-like)
plasma obtained by Perel' and Eliashberg \cite{PE} (see also Refs. \cite{AT3} and
\cite{AMT} for an alternative derivation of this result and its specification
based on the known expression for the bremsstrahlung differential cross
section for high values of energy transfer and $\omega\gg\left(  \beta\hbar\right)  ^{-1}$
\cite{LLVIII}):%
\begin{equation}
\operatorname{Im}\epsilon\left(  k,\omega\gg\left(  \beta\hbar\right)
^{-1}\right)  \simeq\frac{4\pi A}{\omega^{9/2}}, \label{PEas}%
\end{equation}

where%
\begin{equation}
A=\frac{2^{5/2}\pi}{3}\frac{n^{2}e^{6}}{\left(  \hbar m\right)  ^{3/2}}.
\label{Acte}%
\end{equation}

Result (\ref{PEas}) also implies that higher even-order frequency moments, $C_{2l}\left(  k\right)  ,$
$l\geq3,$ diverge.

To take into account all convergent sum rules (power frequency moments) and
the exact asymptotic relation (\ref{PEas}) we apply the Nevanlinna formula
(\ref{eps}) with the following interpolation model expression for the Nevanlinna parameter
function \footnote{This form of the Nevanlinna function follows from the Riesz-Herglotz integral representation (\ref{qg}) with $dg\left( t\right) \sim \sqrt{|t|}dt$.}
\begin{equation}
Q\left(  k,z \right)  =B\left(  k\right)  \sqrt{z }\left(  1 + i \right)+ i h\left(  k\right)\equiv Q_{1}\left(
k,z \right)  +iQ_{2}\left(  k, z \right)  , \label{qB}%
\end{equation}
with%
\[
B\left(  k\right)  =\frac{4\pi A}{\omega_{p}^{2}\left(  \omega_{2}^{2}%
-\omega_{1}^{2}\right)  } \geq 0,
\]
since by virtue of the Cauchy-Schwarz inequality $\omega_{2}^{2}-\omega
_{1}^{2}\geq 0$. Observe that
\[
B\left(  k\right)  \sqrt{\omega_{p}}=\frac{\sqrt{2}}{3^{5/4}}\frac{r_{s}%
^{3/4}\omega_{p}^{3}}{\left(  \omega_{2}^{2}-\omega_{1}^{2}\right)  }%
\]
so that
\[
\frac{Q\left(  k, z \right)  }{\omega_{p}}=\frac{\sqrt{2}}{3^{5/4}}%
\frac{r_{s}^{3/4}\omega_{p}^{2}}{\omega_{2}^{2}-\omega_{1}^{2}}\sqrt
{\frac{ z  }{\omega_{p}} }\left(  1 + i \right)  +i%
\frac{\omega_{2}^{2}-\omega_{1}^{2}}{3\pi\Gamma\omega_{1}^{4}}\frac{\omega
_{p}^{2}q^{2}}{s(q,0)} ,
\]

\[
s(q,0)=\frac{\omega_{p}}{n}S_{zz}\left(  k,0\right)
\]
being the dimensionless charge-charge dynamic structure factor at $\omega=0$
and $q=ka$, the values of $s(q,0)$ are provided in Table IV of Ref. \cite{H2}.

The expression for the (inverse) dielectric function and, hence,
for the dynamic structure factor (\ref{Szz}) with the Nevanlinna parameter
function determined in (\ref{qB}), leads to the correct static value
$S_{zz}(k,0)$ of the dynamic structure factor, satisfies all three sum rules
(\ref{Cnu}), and also satisfies the exact relation (\ref{PEas}).

The range of frequencies studied in Ref. \cite{H2} was about $\left(  0,2\omega
_{p}\right)  $, so that no data was obtained for the frequencies which satisfy
the condition $\omega\gg\left(  \beta\hbar\right)  ^{-1}$ or, equivalently,
\[
\frac{\omega}{\omega_{p}}\gg\sqrt{\frac{r_{s}}{3\Gamma^{2}}}.
\]
On the other hand, in spite of the classical approximations used here for comparison with Ref. \cite{H2}, as we have discussed previously, the system we consider possesses an inherent quantum mechanical nature, and we may presume that the asymptotic form (\ref{PEas}) is
applicable to it.

By inserting Eq. (\ref{qB}) into (\ref{Szz}) we get
\begin{equation}
S_{zz}\left(  k,\omega\right)  =\frac{\omega_{p}^{2}}{\pi\beta\phi\left(
k\right)  }\frac{\left[  \omega_{2}^{2}\left(  k\right)  -\omega_{1}%
^{2}\left(  k\right)  \right]  Q_{2}\left(  k,\omega\right)  }{\left\vert
\omega\left[  \omega^{2}-\omega_{2}^{2}\left(  k\right)  \right]  +Q\left(
k,\omega\right)  \left[  \omega^{2}-\omega_{1}^{2}\left(  k\right)  \right]
\right\vert ^{2}} , \label{Szzkwq}%
\end{equation}
which reduces the determination of the dynamic structure factor to the knowledge of
the static characteristics - the frequency moments $C_{\nu}\left(  k\right)
,$ $\nu=0,2,4$, see Sect. \ref{M}$.$

The latter were calculated back in 1993 in Ref. \cite{ATMV} in terms of the static structure
factors of both system species, $S_{ab}\left( k \right)  $, beyond the
random-phase approximation with the inclusion of both electronic and ionic
local-field corrections (STLC). The electronic STLC was modelled \cite{T,Dj} as an interpolation of the Geldart-Vosko form:
\begin{equation}
G_{e}(k)=\frac{k^{2}}{c_{1} k_{s}^{2}+c_{2} k^{2}}, \label{G1}
\end{equation}
where $k_{s}^{-1}$ is the screening length of the electronic subsystem, treated as a one-component plasma (EOCP), an appropiate interpolation between the Debye and Thomas-Fermi radia (for details see Ref. \cite{ATMV}):
\begin{equation}
\epsilon_{e}\left(k \right) = 1+k_{s}^{2}/k^{2} . \label{G11}
\end{equation}
The interpolation parameters $c_{1}$ and $c_{2}$ were chosen in Ref. \cite{Dj} to satisfy the compressibility sum rule for the classical EOCP \cite{T,Dj} with the equation of state taken from the MD numerical simulations available at that time \cite{SDS}, and the Kimball cusp-condition \cite{K,Ich2}:
\begin{equation}
\lim_{k\to \infty} G_{e}\left(k \right)= 1- g_{e}\left(0 \right). \label{G2}
\end{equation}

The ionic STLC,
\begin{equation}
G_{i}(k)=\frac{k^{2}}{c_{1} k_{s}^{2}+c_{2} k^{2}\epsilon_{e}\left( k\right)}, \label{G3}
\end{equation}
and the structural characteristics, i.e., the zero-separation value of the electronic radial distribution function, $g_{e}(0)$, and the static structure factors, $S_{ab}(k)$, were determined in Ref. \cite{ATMV} according to the Ichimaru algorithm \cite{Ich} in a self-consistent way in terms of the EOCP static structure factor $S_{e}(k)$. To take into account the quantum mechanical corrections, $S_{e}(k)$ was computed in Ref. \cite{ATMV} using the temperature Green's function technique by a regularized summation over the Matsubara frequencies.

To avoid all these cumbersome computations and, more important, to improve the agreement with the simulation data of Ref. \cite{H2} on the dynamic characteristics of dense two-component plasmas, the plasma static characteristics have been calculated in the present work by the HNC method with the Deutsch effective potential, as it was done in Ref. \cite{H2}.

\subsection{The moments \label{M}}

The explicit form of the power moments $C_{\nu}\left(  k\right)  ,$
$\nu=0,2,4$, for the bare Coulomb potential is known since long \cite{AT} (for
details see Ref. \cite{AMT}):
\begin{equation}
C_{0}\left(  k\right)  =\left[  1-\epsilon^{-1}\left(  k,0\right)  \right]
\text{ },\quad C_{2}\left(  k\right)  =\omega_{p}^{2}\text{ },\quad
C_{4}\left(  k\right)  =\omega_{p}^{4}\left[  1+W\left(  k\right)  \right] . \label{CC}%
\end{equation}
The moment $C_{0}\left(  k\right)  $, as it was already commented, is related
to the static dielectric function of the system, the second moment is actually
the $f$-sum rule, which is independent of the system interactions. The correction
in the fourth moment contains different contributions:
\[
W\left(  k\right)  =K\left(  k\right)  +U\left(  k\right)  +H .
\]%
The first contribution stems from the kinetic term of the system Hamiltonian. In the classical case, this coincides with the known Vlasov contribution to the dispersion relation:
\begin{equation}
K(k)= 3 \frac{k^{2}}{k_{D}^{2}}, \label{Kclass}
\end{equation}%
$k_{D}^{2}  =4\pi ne^{2}\beta$ being the square of the Debye wavenumber. Here, as in (\ref{CC}), we only account for the electronic subsystem, due to the large asymmetry between the masses of electrons and ions. We use expression (\ref{Kclass}) for comparison with the results of Ref. \cite{H2}. Nonetheless, due to the quantal nature of our system, it would be interesting to estimate quantitatively how the degeneracy would affect the dispersion law and the dynamic structure factor of the system through this kinetic contribution. In the quantum mechanical case, it can be recast as
\begin{equation}
K(k)= \frac{\left\langle v_{e}^{2} \right\rangle  k^{2}}{\omega_{p}^{2}} +  \left( \frac{\hbar}{2m} \right)^{2}  \frac{k^{4}}{\omega_{p}^{2}}, \label{Kdeg}
\end{equation}%
where the average of the square of the electron velocity is expressed as
\begin{equation*}
\left\langle v_{e}^{2} \right\rangle = 3 \frac{\theta}{m \beta} F_{3/2}(\eta),
\end{equation*}
\begin{equation*}
F_{\nu}(\eta) = \int_{0}^{\infty} \frac{x^{\nu}} {\exp\left( x- \eta\right) +1} dx
\end{equation*}
being the order-$\nu$ Fermi integral, and $\eta=\beta \mu$ the dimensionless chemical potential of the electronic subsystem, which should be determined by the normalization condition
\[
F_{1/2}(\eta) = \frac{2}{3} \theta^{-3/2}.
\]

The last two terms in the fourth moment correction term stem from the interaction contribution to the system Hamiltonian and are, therefore, dependent on the potential used. For the bare Coulomb potential we write:
\begin{equation*}
U\left(  k\right)   =\frac{1}{2\pi^{2}n}%
{\displaystyle\int_{0}^{\infty}}
p^{2}\left(  S_{ee}\left(  p\right)  -1\right)  f\left(  p,k\right)
dp,
\end{equation*}
\begin{equation*}
H   =\frac{1}{3}h_{ei}\left(  0\right)  =\frac{1}{3}\left(  g_{ei}\left(
0\right)  -1\right)  =\frac{1}{6\pi^{2}n}%
{\displaystyle\int_{0}^{\infty}}
p^{2}S_{ei}\left(  p\right)  dp ,
\end{equation*}
where we have introduced
\begin{equation*}
f\left(  p,k\right)    =\frac{5}{12}-\frac{p^{2}}{4k^{2}}+\frac{\left(
k^{2}-p^{2}\right)  ^{2}}{8pk^{3}}\ln\left\vert \frac{p+k}{p-k}\right\vert .
\end{equation*}

But for the sake of a better comparison with the MD results of Ref. \cite{H2}, we
recalculated here these moments using the model Hamiltonian with the Coulomb
potential substituted by the model Deutsch effective potential. Then, the moment
$C_{2}\left(  k\right)  $ (the $f$\textit{-}sum rule) remains intact, the
moment $C_{0}\left(  k\right)  $ changes together with the model system static dielectric function, and
instead of $C_{4}\left(  k\right)  $ we have:
\begin{equation}
\widetilde{C}_{4}\left(  k\right)  =\omega_{p}^{4}\left[  1+\widetilde
{W}\left(  k\right)  \right]  , \label{c4t}%
\end{equation}
where the \textquotedblright model\textquotedblright\ $\widetilde{W}\left(
k\right)  $ has the same kinetic contribution $K\left(  k\right)  $ (either classical or degenerate), but the
interaction contributions are substituted by
\begin{equation}
\widetilde{U}\left(  k\right)  =\frac{1}{2\pi^{2}n}%
{\displaystyle\int_{0}^{\infty}}
p^{2}\left(  \widetilde{S}_{ee}\left(  p\right)  -1\right)  \widetilde
{f}\left(  p,k\right)  dp \label{ut}%
\end{equation}
and
\begin{equation}
\widetilde{H}=\frac{\kappa_{ei}^{2}}{6\pi^{2}n}%
{\displaystyle\int_{0}^{\infty}}
\frac{p^{2}\widetilde{S}_{ei}\left(  p\right)  }{p^{2}+\kappa_{ei}^{2}}%
dp . \label{Ht}%
\end{equation}
Here%
\begin{eqnarray*}
\widetilde{f}\left(  p,k\right)  &=& \frac{\kappa_{ee}^{2}}{4k^{2}}+\frac{\left(
k^{2}-p^{2}\right)  ^{2}}{8pk^{3}}\ln\left\vert \frac{p+k}{p-k}\right\vert -
\\
&-&   \frac{\left(  p^{2}+\kappa_{ee}^{2}-k^{2}\right)  ^{2}}{16pk^{3}}\ln
\frac{\left(  p+k\right)  ^{2}+\kappa_{ee}^{2}}{\left(  p-k\right)
^{2}+\kappa_{ee}^{2}}-\frac{\kappa_{ee}^{2}/3}{p^{2}+\kappa_{ee}^{2}} .
\end{eqnarray*}

In addition, the model partial static structure factors $\widetilde{S}_{ab}\left(
k\right)  $ have been computed in the hypernetted-chain approximation using the
Deutsch effective potential. This closes the algorithm of calculation of the static and dynamic characteristics of the system.

Some numerical results are discussed in the following Section.

\section{Numerical results}

\subsection{Static characteristics \label{Stchar}}

As it was mentioned, we calculated the static structure factors and the radial
distribution functions in the hypernetted-chains approximation using the
Deutsch effective potential (\ref{Dpot}), just as it was done in Ref. \cite{H2}. 

Some data on the partial static structure factors are presented in Tables \ref{table1}-\ref{table3}.
It is not surprising that the agreement we obtain with the values of the
static characteristics given in Table \ref{table5} of Ref. \cite{H2}, is within the
computational error. We also add the corresponding values of the charge-charge
static structure factor $S_{zz}(k)$ calculated by the following relation:
\begin{equation}
S_{zz}(k)=S_{ii}(k)+S_{ee}(k)-2S_{ie}(k)\label{szzkstatic}%
\end{equation}
and check the zero-order frequency moment of the dynamic factor (\ref{Szzkwq}), %
\begin{equation}
S_{zz}(k)=\frac{1}{n}%
{\displaystyle\int_{-\infty}^{\infty}}
S_{zz}(k,\omega)d\omega ;\label{szzkdynamic}%
\end{equation}
to find an agreement to the fourth decimal digit.

The other two dimensionless even-order power moments of the dynamic structure
factor (\ref{Szzkwq}) are defined as
\begin{equation}
S_{\nu}\left(  k\right)  =\frac{1}{n\omega_{p}^{\nu}}%
{\displaystyle\int_{-\infty}^{\infty}}
\omega^{\nu}S_{zz}(k,\omega)d\omega ,\quad\nu =2,4, \label{snd}%
\end{equation}
the latter being provided in Table \ref{table4}. By virtue of the classical version of the FDT used here, the odd-order moments vanish due to the symmetry of
(\ref{Szzkwq}).

These values are then used to determine the characteristic frequencies $\omega_{1}\left(  k\right)  = \sqrt{ C_{2}\left(  k\right)  /\widetilde{C}_{0}\left(  k\right) }  $ and $\omega_{2}\left(
k\right)  =\sqrt{\widetilde{C}_{4}\left(  k\right)  /C_{2}\left(  k\right)  }$ which virtually coincide with their values calculated
from the formulas (\ref{c4t}), (\ref{ut}), and (\ref{Ht}), but now differ
significantly from the values given in Table VI of Ref. \cite{ATMV}.

Notice that due to (\ref{eps}), (\ref{compress}) and the FDT, the static inverse dielectric function and the moment $\widetilde{C}
_{0}\left(  k\right)  $ are directly related to the charge-charge static
structure factor (\ref{szzkdynamic}):
\begin{equation}
S_{zz}\left(  k\right)  =-\frac{k^{2}}{k_{D}^{2}}P.V.%
{\displaystyle\int_{-\infty}^{\infty}}
\operatorname{Im}\epsilon^{-1}\left(  k,\omega\right)  \frac{d\omega}{\pi \omega
}=\frac{k^{2}}{k_{D}^{2}}\left(  1-\operatorname{Re}\epsilon^{-1}\left(
k,0\right)  \right)  ,  \label{SzzFDT}
\end{equation}
Thus the moment $\widetilde{C}_{0}\left(  k\right)  $ was estimated as
\begin{equation}
\widetilde{C}_{0}\left(  k\right)  =\frac{k_{D}^{2}}{k^{2}}S_{zz}\left(
k\right)  \label{C0FDT}
\end{equation}
with the static structure factor $S_{zz}\left(  k\right)  $ also calculated in the
hypernetted-chain approximation using the Deutsch effective potential (\ref{Dpot}). 

Further, we display our results on the values of the partial radial
distribution functions at zero separation, $g_{ee}(0)$ and $g_{ie}(0)$,
computed using the effective potential (\ref{Dpot}) and also taking into account (within the HNC scheme) the symmetry
effects in the electron-electron exchange contribution to the
effective potential (while leaving other components unchanged):
\begin{equation}
\varphi_{ee}\left(  r\right)  =\frac{e^{2}}{r}\left(  1-\exp\left(
-\kappa_{ee}r\right)  \right)  +\frac{\ln2}{\beta}\exp\left(  -\frac
{r^{2}\kappa_{ee}^{2}}{\pi\ln2}\right)  ; \label{Dx}%
\end{equation}
for comparison we present also the values of $g_{ie}(0)$ calculated
analytically in Ref. \cite{TOR}, see Table \ref{table5}.

\begin{figure*}
\centering{
\includegraphics[width=.45\textwidth]{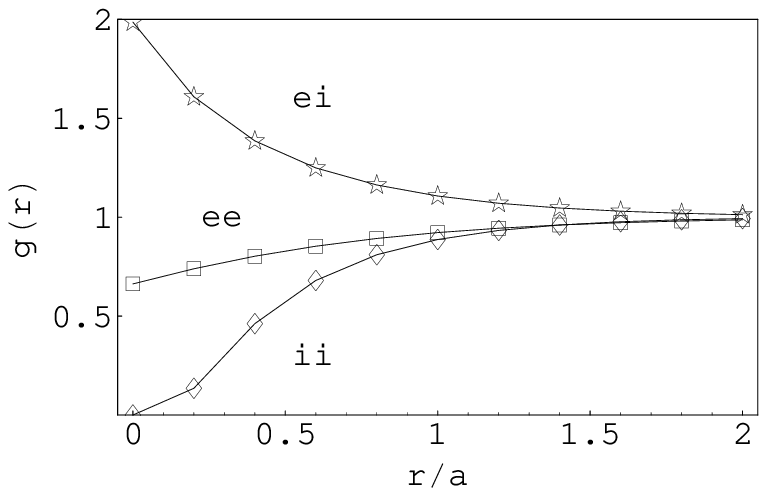}
}
\caption{}
\label{fig:pcf}
\end{figure*}

In addition, we display the graphs for the three partial radial distribution
functions for the conditions $\Gamma=0.5$ and $r_{s}=0.4$, calculated with the
potential (\ref{Dpot}), Fig. \ref{fig:pcf}. The curves in Fig. \ref{fig:pcf} are virtually
indistinguishable from those of Fig. 2 of Ref. \cite{H2}.

\subsection{Dynamic characteristics}

Results on the dynamic structure factor itself are broadly presented in Figs. \ref{fig:szz-g05-r04}-\ref{fig:szz-g20-r10}.

The agreement between these graphs and the corresponding MD results of
\cite{H2} is quantitatively good, and
it is better than that achieved in Ref. \cite{ATMV}. The introduction of the
non-constant Nevanlinna parameter function (\ref{qB}) not only permits to obtain better
agreement in the position of the Langmuir peaks, but also leads to the adequate broadening (damping) of the Langmuir mode. 

In this sense, it is interesting to calculate the complex solution for the dispersion equation $\epsilon\left (k,z \right )=0$ explicitly in order to determine quantitatively the damping of the collective mode. From Eq. (\ref{epsx}) we get the equation
\begin{equation}
z\left( z^{2} - \omega_{2}^{2}\right) + Q\left ( z^{2}- \omega_{1}^{2} \right ) = 0 . \label{dispersion}
\end{equation}
Due to the fact that the function $\epsilon^{-1}(k,z)$ must be analytic in the upper half-plane, the solution of the dispersion equation, $z(k)$, must possess a non-positive imaginary part, i.e., if $z(k) = {\rm Re}z(k) + i {\rm Im}z(k) $, then ${\rm Im}z \leq 0$. In particular, it is clear that for the Feynman-like approximation $Q(h,z)=i0^{+}$ we get for the (shifted) collective excitation the value of $\omega_{2}$. However, for the model function (\ref{qB}), it is shown in Figs. \ref{fig:dispersion-class} and \ref{fig:dispersion-deg} that ${\rm Re}z < \omega_{2}$, while the damping becomes more notorious.

We observe that in all cases, except for $\Gamma=2$ and the lowest values of $q=ka$, the effects of degeneracy produce stronger positive dispersion, i.e., the Langmuir mode frequency become higher than $\omega_{p}$. In addition, degeneracy or quantum mechanical characteristics of the effective interaction produce also stronger damping of the mode.

Indeed, at higher frequencies, we deal with shorter distances or shorter times, where the non-Coulomb nature of the effective potential and the wave nature of electrons become more pronounced.

Notice that these effects are reflected in the MD calculations using the Deutsch effective potential and are described in our calculations.

\section{Conclusions}

The agreement with the MD data on the dynamic structure factor and other
dynamic characteristics of the model system, like the Langmuir collective mode
dispersion, is improved with respect to the results obtained in Ref. \cite{ATMV},
and, simultaneously, important statistical characteristics of the model TCP in
concern, like the probability to find both electron and ion at one point
$\left[  g_{ei}\left(  0\right)  \right]  $, are determined, the effects of direct and exchange interactions being compared.  

The importance of (\ref{Kdeg}) and the applicability (\ref{qB}) are validated.

Since the static characteristics of the system (the static structure factors and the radial distribution functions) were computed with high precision in the way employed in Ref. \cite{H2}, the reliability of the calculation of the characteristic frequencies $\omega_{1}(k)$ and $\omega_{2}(k)$ is improved. These quantities are essential to estimate the position and damping of the plasma collective mode.

Another key ingredient for the good quantitative agreement achieved with the results of Ref. \cite{H2} with respect to the dispersion and decay of the Langmuir mode, is the introduction of a non-constant Nevanlinna parameter function accounting for the exact high-frequency asymptotic form of the system dielectric function. Still, further specification of this dynamic characteristic similar to the dynamic local-field correction in a TCP might be needed. 

We observed that for the Nevanlinna parameter function $Q=Q(k,z)$ considered here, the value of the Langmuir frequency shifts from $\omega_{2}(k)$ closer to the plasma frequency, the effect of which might be considered correct from the experimental point of view. 

Further extension of the present mixed approach to more recent data for the
model systems described by other effective potentials \cite{ZP,M} is planned.

\section{Acknowledgments}

The authors acknowledge the financial support of the \textit{Fundaci\'{o}n
Universidad Internacional de Valencia} and the \textit{al-Farabi Kazakh National University}.


\newpage

{\bf Table Captions}\newline

Table I. Partial static structure factors at $\Gamma=0.5,$ $r_{s}=0.4$ with the Deutsch effective potential without exchange (\ref{Dpot}).\newline

Table II. Same as in Table \ref{table1}, but for $\Gamma=0.5$, $r_{s}=1$.\newline

Table III. Same as in Table \ref{table1}, but for $\Gamma=2$, $r_{s}=1$.\newline

Table IV. The fourth dimensionless power moment of $S_{zz}\left( k,\omega\right)$, according to (\ref{snd}), with the Deutsch effective potential without exchange (\ref{Dpot}). In the classical case, the kinetic contribution is approximated by the Vlasov term (\ref{Kclass}), whereas in the quantal case expression (\ref{Kdeg}) is used.\newline

Table V. Zero-separation values of the partial radial distribution functions
compared to the results of Ref. \cite{TOR}.\newline

\newpage

\begin{table*}
\caption{\label{table1}  }
\begin{ruledtabular}
\begin{tabular}{ccccc}
$q=ka$ & $S_{ii}(k)$ & $S_{ie}(k)$ & $S_{ee}(k)$ & $S_{zz}(k)$\footnote{Calculated from (\ref{szzkstatic}) or (\ref{szzkdynamic}).} \\ \hline
0.767 & 0.5804 & 0.4387 & 0.6590 & 0.3620\\
1.074 & 0.6257 & 0.3600 & 0.7391 & 0.6448\\
1.381 & 0.6824 & 0.2813 & 0.8118 & 0.9316\\
1.534 \  & 0.7118 & 0.2455 & 0.8425 & 1.0634
\end{tabular}
\end{ruledtabular}
\end{table*}

\vspace*{2cm}

\begin{table*}
\caption{\label{table2}  }
\begin{ruledtabular}
\begin{tabular}{ccccc}
$q=ka$ & $S_{ii}(k)$ & $S_{ie}(k)$ & $S_{ee}(k)$ & $S_{zz}(k)$\footnote{Calculated from (\ref{szzkstatic}) or (\ref{szzkdynamic}).} \\ \hline
0.767 & 0.6160 & 0.4606 & 0.6470 & 0.3418\\
1.074 & 0.6663 & 0.3943 & 0.7144 & 0.5922\\
1.381 & 0.7192 & 0.3275 & 0.7769 & 0.8412\\
1.534 & 0.7447 & 0.2952 & 0.8081 & 0.9624
\end{tabular}
\end{ruledtabular}
\end{table*}

\vspace*{2cm}

\begin{table*}
\caption{\label{table3}  }
\begin{ruledtabular}
\begin{tabular}{ccccc}
$q=ka$ & $S_{ii}(k)$ & $S_{ie}(k)$ & $S_{ee}(k)$ & $S_{zz}(k)$\footnote{Calculated from (\ref{szzkstatic}) or (\ref{szzkdynamic}).} \\ \hline
0.767 & 0.5642 & 0.5821 & 0.7197 & 0.1198\\
1.074 & 0.5133 & 0.5001 & 0.7385 & 0.2516\\
1.381 & 0.5067 & 0.4275 & 0.7769 & 0.4286\\
1.534 & 0.5174 & 0.3940 & 0.7993 & 0.5288
\end{tabular}
\end{ruledtabular}
\end{table*}

\vspace*{2cm}

\begin{table*}
\caption{\label{table4}  }
\begin{ruledtabular}
\begin{tabular}{ddddddd}
 &\multicolumn{3}{c}{Classical}&\multicolumn{3}{c}{Quantal}\\
 & \multicolumn{1}{r}{$\Gamma=0.5$} & \multicolumn{1}{r}{$\Gamma=0.5$} & \multicolumn{1}{r}{$\Gamma=2.0$} & \multicolumn{1}{r}{$\Gamma=0.5$} & \multicolumn{1}{r}{$\Gamma=0.5$} & \multicolumn{1}{r}{$\Gamma=2.0$} \\
\multicolumn{1}{r}{$q=ka$} & \multicolumn{1}{r}{$r_{s}=0.4$} & \multicolumn{1}{r}{$r_{s}=1.0$} & \multicolumn{1}{r}{$r_{s}=1.0$} & \multicolumn{1}{r}{$r_{s}=0.4$} & \multicolumn{1}{r}{$r_{s}=1.0$} & \multicolumn{1}{r}{$r_{s}=1.0$} \\ \hline
0.767 & 0.8845 & 0.9318 & 0.1403 & 1.1175 & 0.9966 & 0.1680\\
1.074 & 2.6028 & 2.6943 & 0.3294 & 3.6028 & 2.9856 & 0.4462\\
1.381 & 6.2193 & 6.3685 & 0.6644 & 9.3337 & 7.3169 & 1.0217\\
1.534 & 9.0728 & 9.2555 & 0.9073 & 14.1572 & 10.8366 & 1.4855
\end{tabular}
\end{ruledtabular}
\end{table*}

\vspace*{2cm}

\begin{table*}
\caption{\label{table5}  }
\begin{ruledtabular}
\begin{tabular}{cccccccc}
$\Gamma$ & $\theta$ & $r_{s}$ & $g_{ee}(0)$\footnotemark[1] & $g_{ie}(0)$\footnotemark[1] & $g_{ee}(0)$\footnotemark[2] & $g_{ie}(0)$\footnotemark[2] & $g_{ie}(0)$\footnotemark[3]\\ \hline
0.1 & 0.1000 & 0.0184 & 0.9598 & 1.0684 & 0.7309 & 1.0474 & 1.0358\\
0.1 & 2.0000 & 0.3683 & 0.7569 & 1.5164 & 0.3868 & 1.5045 & 1.3478\\
0.5 & 0.4344 & 0.4000 & 0.6629 & 1.9865 & 0.3897 & 1.8874 & 1.4412\\
0.5 & 1.0860 & 1.0000 & 0.4707 & 3.5224 & 0.2500 & 3.2600 & 1.8963\\
1.0 & 0.1000 & 0.1842 & 0.7738 & 1.4979 & 0.6159 & 1.4276 & 1.2351
\end{tabular}
\end{ruledtabular}
\footnotetext[1]{With (\ref{Dpot}).}
\footnotetext[2]{With (\ref{Dpot})+(\ref{Dx}).}
\footnotetext[3]{Values from Ref.~\cite{TOR}.}
\end{table*}

\vspace*{10cm}


\newpage

{\bf Figure Captions}\newline

Fig. 1. The radial distribution functions $g_{ab}(r)$, $a,b=e,i$ for $\Gamma=0.5$, $r_{s}=0.4$, calculated with the potential (\ref{Dpot}). Solid lines are included to join the discrete points for a better visualization.\newline

Fig 2. Normalized dynamic structure factor versus frequency as given by expression (\ref{Szzkwq}) at $\Gamma=0.5$, $r_{s}=0.4$. The model parameter function (\ref{qB}) is used. The values of $ka$ are: (1) 0.767, (2) 1.074, (3) 1.381, and (4) 1.534. (a) Classical calculations (solid lines) compared to the results of Ref. \cite{H2} with $ka=0.780$ (diamonds), $ka=1.102$ (triangles), $ka=1.350$ (boxes), and $ka=1.559$ (stars). (b) Comparison between classical (dashed) and quantal calculations (solid). In the classical case, the kinetic contribution is approximated by the Vlasov term (\ref{Kclass}), whereas in the quantal case expression (\ref{Kdeg}) is used.\newline

\begin{figure*}
\centering{
\includegraphics[width=.45\textwidth]{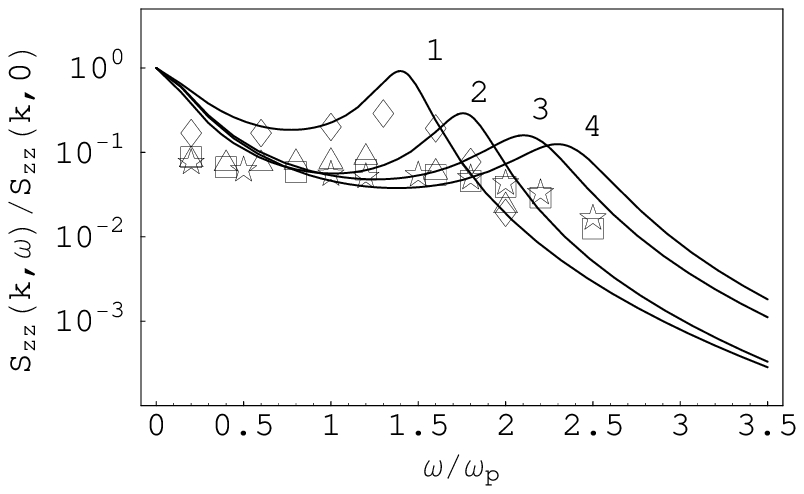}~(a)
\hfil
\includegraphics[width=.45\textwidth]{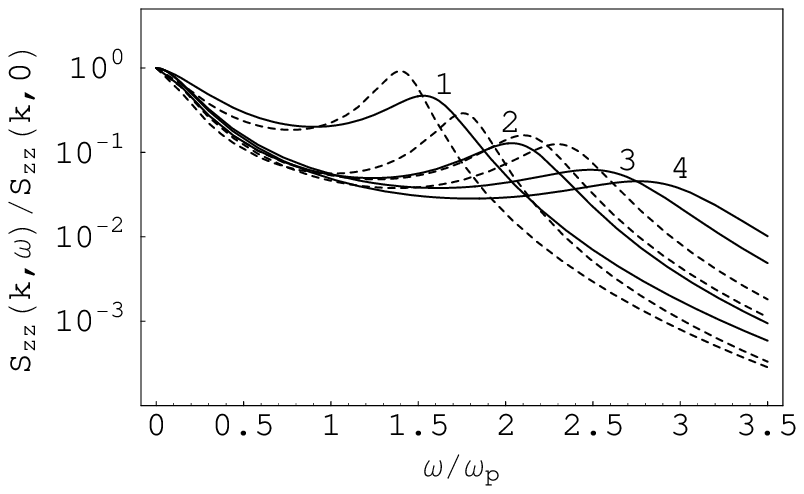}~(b)
}
\caption{}
\label{fig:szz-g05-r04}
\end{figure*}

\begin{figure*}
\centering{
\includegraphics[width=.45\textwidth]{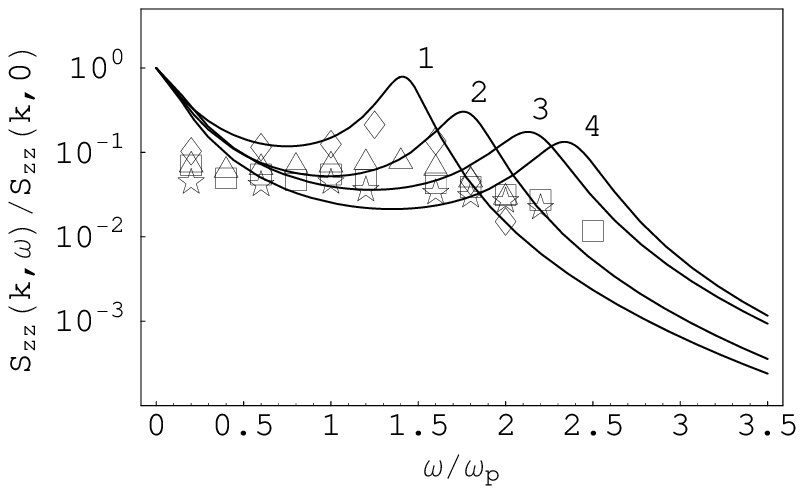}~(a)
\hfil
\includegraphics[width=.45\textwidth]{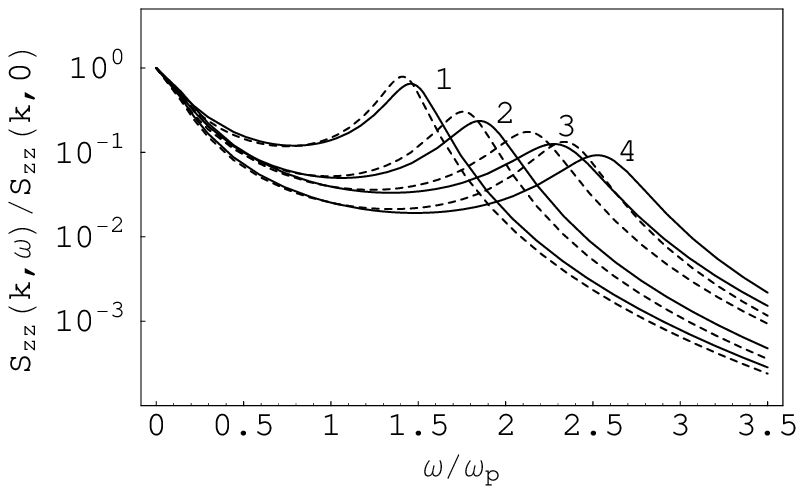}~(b)
}
\caption{}
\label{fig:szz-g05-r10}
\end{figure*}

\begin{figure*}
\centering{
\includegraphics[width=.45\textwidth]{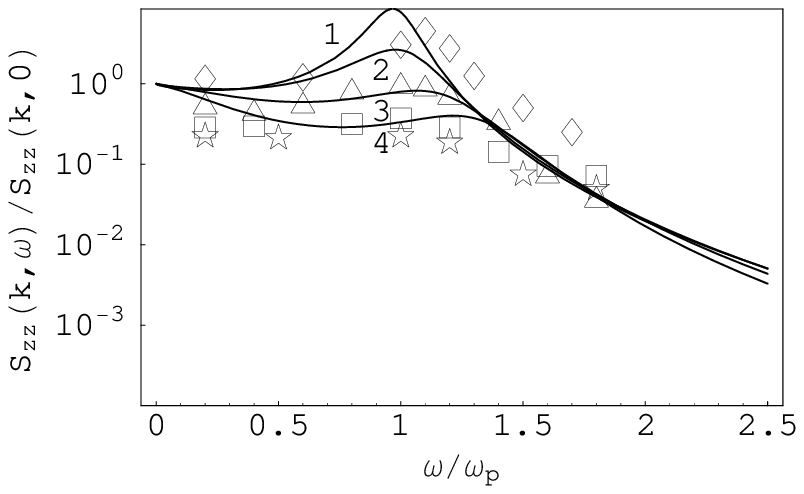}~(a)
\hfil
\includegraphics[width=.45\textwidth]{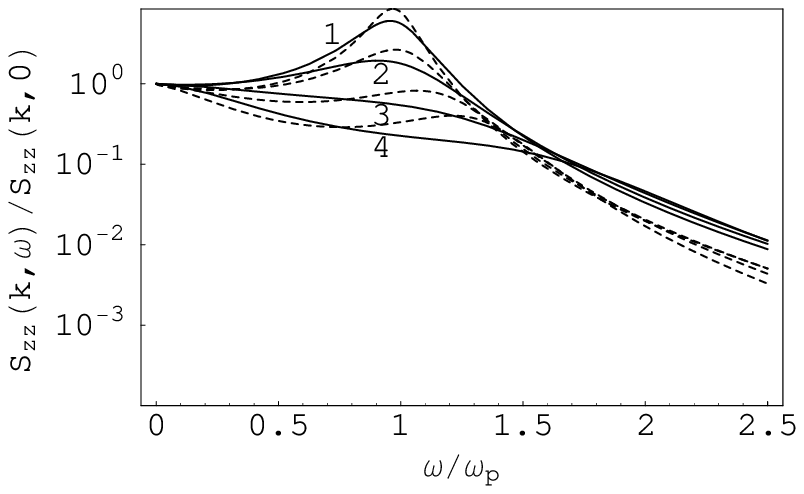}~(b)
}
\caption{}
\label{fig:szz-g20-r10}
\end{figure*}

Fig. 3. Same as in Fig. \ref{fig:szz-g05-r04}, but for $\Gamma=0.5$, $r_{s}=1$.\newline

Fig. 4. Same as in Fig. \ref{fig:szz-g05-r04}, but for $\Gamma=2$, $r_{s}=1$.\newline

Fig. 5. Dispersion, $z(k)  = {\rm Re}z(k) + i {\rm Im}z(k) $, for the collective excitation mode obtained from Eq. (\ref{dispersion}) in the classical approximation. The damped solution stemming from the model function (\ref{qB}) is compared to the undamped solution corresponding to the Feynman-like approximation (\ref{can}). For expression (\ref{qB}): $\Gamma=0.5$, $r_{s}=0.4$ (triangles); $\Gamma=0.5$, $r_{s}=1$ (stars); $\Gamma=2$, $r_{s}=1$ (pentagons). For expression (\ref{can}): $\Gamma=0.5$, $r_{s}=0.4$ (boxes); $\Gamma=0.5$, $r_{s}=1$ (diamonds); $\Gamma=2$, $r_{s}=1$ (circles). Solid lines are included to join the discrete points for a better visualization.\newline

\begin{figure*}
\centering{
\includegraphics[width=.45\textwidth]{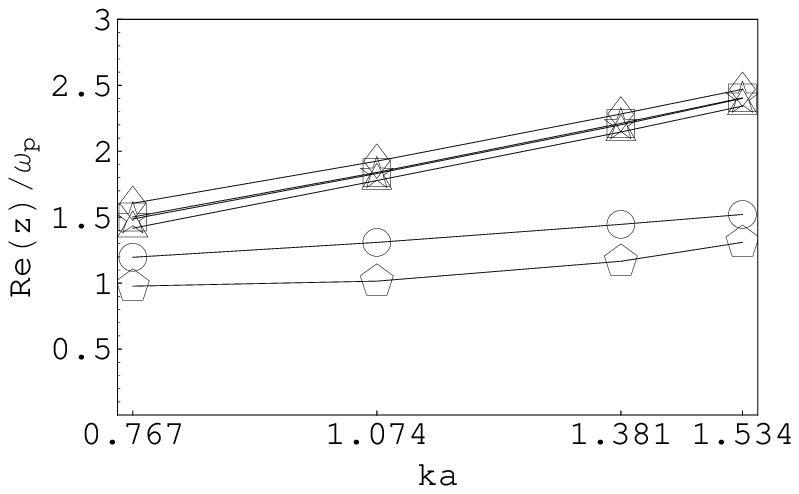}~(a)
\hfil
\includegraphics[width=.45\textwidth]{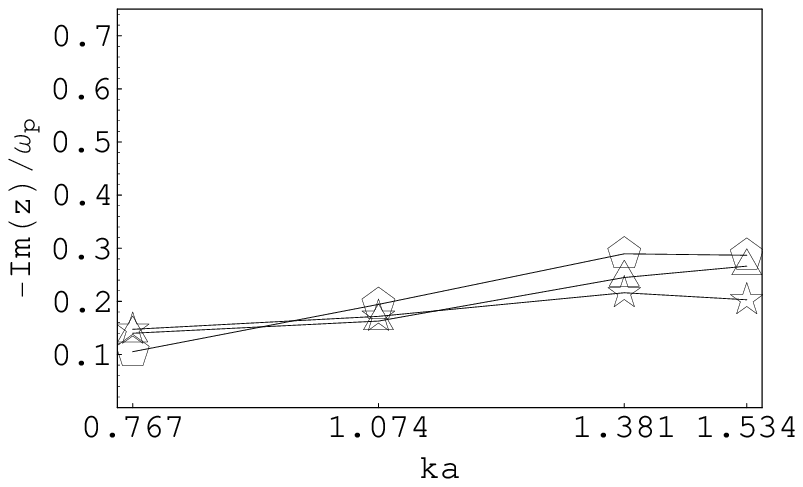}~(b)
}
\caption{}
\label{fig:dispersion-class}
\end{figure*}

\begin{figure*}
\centering{
\includegraphics[width=.45\textwidth]{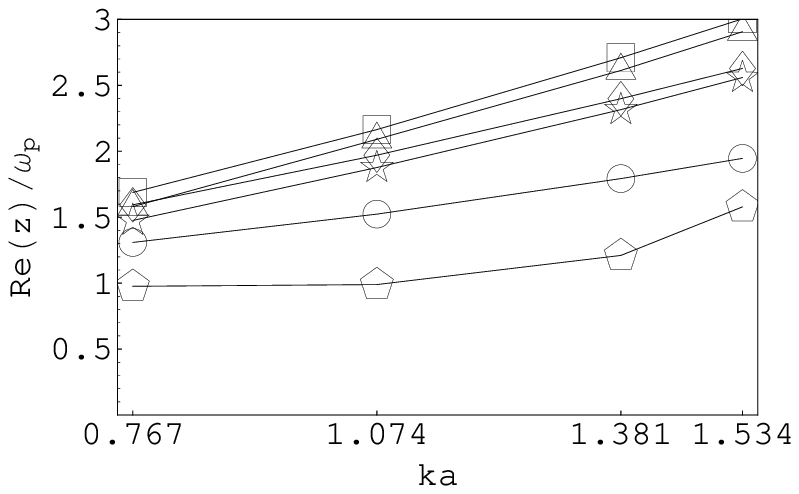}~(a)
\hfil
\includegraphics[width=.45\textwidth]{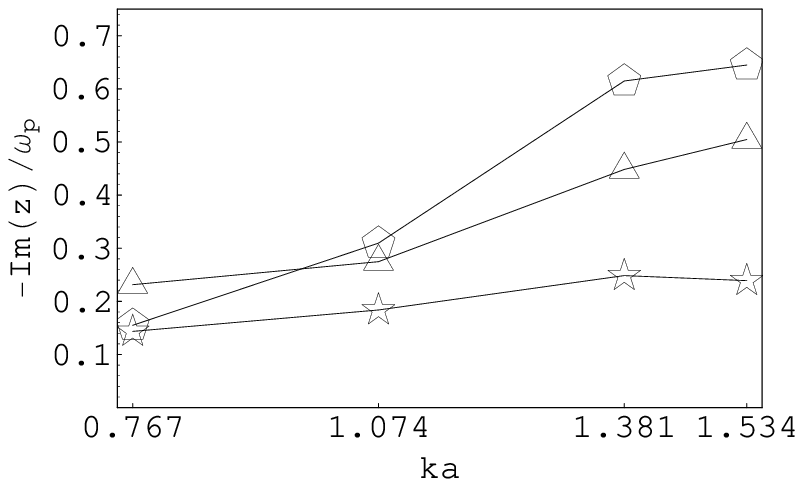}~(b)
}
\caption{}
\label{fig:dispersion-deg}
\end{figure*}

Fig. 6. Same as in Fig. \ref{fig:dispersion-class}, but for the quantal case. The degenerate kinetic term (\ref{Kdeg}) is used, instead of the classical one (\ref{Kclass}).\newline

\newpage

\vspace*{2cm}

\vspace*{2cm}

\vspace*{2cm}

\vspace*{2cm}

\vspace*{2cm}

\end{document}